\documentclass[final,5p,twocolumn]{elsarticle}

\usepackage{hyperref}
\usepackage{multirow}
\usepackage{float}
\usepackage{colortbl}
\usepackage{xcolor}
\usepackage{graphicx}
\usepackage{dcolumn}
\usepackage{bm}
\usepackage{booktabs}
\usepackage{makecell}

\usepackage{amssymb}
\usepackage{amsmath,amsfonts}
\usepackage[utf8] {inputenc}
\usepackage[normalem]{ulem}
\usepackage[T1]{fontenc} 
\usepackage{booktabs}
\usepackage{threeparttable}
\usepackage{tabularx}
\biboptions{sort&compress}

\begin{document}

\begin{frontmatter}

\title{The recurrence of groups inhibits the information spreading under higher‑order interactions}
\author[1]{Liang Yuan}
\affiliation[1]{organization={School of Physics and Electronic Engineering, Jiangsu University},
            city={Zhenjiang},            
            state={Jiangsu},
            postcode={213012}, 
            country={China}}

\author[2]{Jiao Wu}           
\affiliation[2]{organization={School of Mathematical Sciences, Jiangsu University},
	city={Zhenjiang},	
	state={Jiangsu},
	postcode={213012}, 
	country={China}}

\author[1]{Kesheng Xu}
\author[1]{Muhua Zheng\corref{cor1}}
\ead{zhengmuhua163@gmail.com}

\cortext[cor1]{Corresponding author}

\date{\today}

\begin{abstract}
Modeling social systems as networks based on pairwise interactions between individuals offers valuable insights into the mechanisms underlying their dynamics. 
However, the majority of social interactions occur within groups of individuals, characterized by higher-order structures. The mechanisms driving group formation and the impact of higher-order interactions, which arise from group dynamics, on information spreading in face-to-face interaction networks remain insufficiently understood. 
In this study, we examine some representative human face-to-face interaction data and find the recurrent patterns of groups. Moreover, we extend the force-directed motion (FDM) model with the forces derived from similarity distances within a hidden space to reproduce the recurrent group patterns and many key properties of face-to-face interaction networks. Furthermore, we demonstrate that the FDM model effectively predicts information-spreading behaviors under higher-order interactions. Finally, our results reveal that the recurrence of triangular groups inhibits the spread of information in face-to-face interaction networks, and the higher-order interactions will make this phenomenon more pronounced. These findings represent a significant advancement in the understanding of group formation and may open new avenues for research into the effects of group interactions on information propagation processes. 
\end{abstract}

                            

\end{frontmatter}

\section{Introduction}\label{intro}


Comprehending the mechanisms that govern the dynamics
of face-to-face interaction networks is essential for advancing the investigation of spreading processes, such as the transmission of diseases, the word-of-mouth spread of opinions and rumors, social norms contagion, and so on~\cite{barrat2015face,masuda2016guide,holme2015modern}. A typical description of face-to-face interactions is temporal networks formed from a sequence of network snapshots~\cite{holme2012temporal}. In each snapshot, nodes characterize the individuals, and an edge connecting two nodes indicates an interaction. Temporal networks have provided a powerful tool for us to determine many structural and dynamical properties of social interactions~\cite{holme2012temporal,barbosa2018human,starnini2017robust}, such as the distribution of the communication duration and of interconversation times~\cite{isella2011s,cattuto2010dynamics}. Two representative models have been proposed to reproduce quantitatively many essential characteristics of real-world interaction networks. One is the attractiveness model (AM)~\cite{starnini2013modeling,starnini2016model}, in which social attractiveness is the dominant factor in adjusting the motion of individuals. Another model is the force-directed motion (FDM) model~\cite{flores2018similarity}, which posits that individuals inhabit a latent similarity space, where the distances between them in this space generate similarity forces that govern their movements within the physical space.

However, the previous models and empirical analyses of human  
interaction networks mainly focus on pairwise interactions, which do not capture the complicated group interactions composed of three or more individuals (i.e., higher‑order interactions)~\cite{iacopini2019simplicial,matamalas2020abrupt,wang2024correlation,battiston2021physics,battiston2020networks,boccaletti2023structure, starnini2013modeling,starnini2016model,greene2010tracking,flores2018similarity}.
In face-to-face interaction networks, individuals generally form groups with close social circles and to the benefit of information communication. For example, different numbers of individuals converse in meetings or during social gatherings.
Group interactions can typically be described using simplicial complexes~\cite{iacopini2019simplicial}. A group $g =[v_0, v_1, ...,v_k]$ consisting of 
$k + 1$ nodes is referred to as a $k$-simplex. 
It is a higher-dimensional generalization of edges, faces, and volumes in a topological space.  
For example, a group of three individuals can be described by $2$-simplex in a full triangle $[v_0,v_1,v_2]$, along with the corresponding edge set $[v_0,v_1]$, $[v_0,v_2]$, $[v_1,v_2]$.  
Research has shown that many collective dynamical behaviors are significantly affected by higher‑order interactions in simplicial complexes, such as epidemic spreading~\cite{iacopini2019simplicial,de2020social}, diffusion~\cite{schaub2020random,carletti2020random},  ­synchronization~\cite{skardal2019abrupt,millan2020explosive,lucas2020multiorder}, social dynamics~\cite{neuhauser2020multibody,sahasrabuddhe2021modelling,jusup2022social}, and ­games~\cite{jusup2022social,alvarez2021evolutionary}. 
More importantly, recent advances have shown that incorporating higher-order architecture can greatly enhance our understanding and predictive ability of their dynamics, such as in the signal propagation~\cite{JI20231}, the evolution of honesty~\cite{kumar2021evolution}, and cooperation
in evolutionary dynamics etc.~\cite{alvarez2021evolutionary} (see the review Refs.~\cite{majhi2022dynamics,boccaletti2023structure} for details). In this sense, a complete understanding of higher‑order interactions in social interaction networks has become a compulsory assignment~\cite{dunbar2018anatomy,latora2017complex,vespignani2018twenty}.

As we know, several pioneering studies have drawn attention to the temporal dynamics of group interactions~\cite{iacopini2024temporal,cencetti2021temporal,wang2023evolutionary,moinet2018effect,gallo2024higher}. For instance, Iacopo \textit{et al.} explored the temporal dynamics of groups via real-world traces of social interactions and focused on how groups form, transition, and dissolve over time~\cite{iacopini2024temporal}. Moreover, Gallo \textit{et al.} 
proposed a model with social attractiveness to reproduce  different properties of groups in face-to-face interactions
and the homophilic patterns at the level of higher-order
interactions~\cite{gallo2024higher}.

Despite fruitful efforts that have recently paid attention to higher-order networks, the mechanisms that force the formation of groups in face-to-face interaction networks remain poorly understood. In addition, the effect of higher‑order interactions resulting from groups in temporal networks on the information-spreading process is yet to be fully explored. In this work, we obtain some representative human face-to-face interaction network data~\cite{sociopatterns} and find the recurrent patterns of groups by analyzing these data. 
These findings stimulate scientific inquiry and drive further investigation into the mechanisms underlying the recurrence of groups within real networks, as well as the impact of group recurrence on information diffusion processes.
Interestingly, we find a natural explanation by extending the FDM model~\cite{flores2018similarity} that similarity forces as a mechanism responsible for individuals' motions. In addition, this model can capture the observed recurrent patterns of groups and many crucial features of real temporal networks.
Finally, we reveal that the recurrence of groups suppresses the information spreading in both real and artificial temporal networks.

The arrangement of this article is as follows. We provide a brief overview of this study's human face-to-face interaction datasets in Section~\ref{datasets}. Section~\ref{Model} presents the network models employed to replicate the recurrent group patterns and various characteristics of real networks. In Section~\ref{SIS}, we introduce a higher-order Susceptible-Infected-Susceptible (SIS) spreading model~\cite{iacopini2019simplicial, wang2024correlation,  matamalas2020abrupt} applied to both real and synthetic temporal networks, to investigate the influence of group recurrence on spreading dynamics. Additionally, we present a simplified theoretical framework utilizing the microscopic Markov chain approach (MMCA) to examine the outcomes of the numerical simulations. Section~\ref{results} presents the results of the simulation and MMCA on real and synthetic temporal networks. In the end, Section~\ref{Conclusion} provides a conclusion.

\section{Human face-to-face interaction datasets}\label{datasets}
We obtain the human face-to-face interaction datasets from Ref.~\cite{flores2018similarity}. These datasets were collected in the SocioPatterns collaboration project~\cite{sociopatterns,vanhems2013estimating,mastrandrea2015contact,pintonhigh} using individuals' radio-frequency identification tags with time slots at 20-second intervals. An interaction was recorded only when two tags were within a $1$ to $1.5$ meters range. The description of considered networks in this study is as follows:  

\textbf{(i) Primary school:} This dataset was gathered from a primary school in Lyon, France. The data comprises 2 periods with $1555$ and $1545$ time slots, respectively. The total duration is $3100$ slots involving 242 nodes. 
We use the first period with time slots $T=1555$ to estimate the parameters of the models and validate the models with the rest of the data.

\textbf{(ii) High school:} This dataset was collected from a high school in Marseille, France. This data consists of 5 periods. The first period lasts $899$ time slots, while the subsequent four periods each last $1619$ time slots, resulting in a total of 7375 time slots involving 327 nodes.  
We use the first three activity periods, comprising $T=4137$ time slots, to estimate the parameters of the models and validate the models with the rest of the data.

\textbf{(iii) Hospital:} This interaction data was gathered from a hospital ward in Lyon, France. The data includes 4 periods with $4400$ time slots involving 70 nodes. We use the first two periods, comprising $T=2200$ time slots, to estimate the parameters of the models and validate the models with the rest of the data.

\textbf{(iv) Conference:}  This dataset was gathered from a scientific conference in Turin, Italy. The data includes 3 periods with $7030$ time slots involving 113 nodes. We estimate the parameters of the models using the time slots $T=2874$ from the first period and validate the models with the rest of the data.

\begin{figure*}[!ht]
	\centering
	\includegraphics[width=1.0\textwidth]{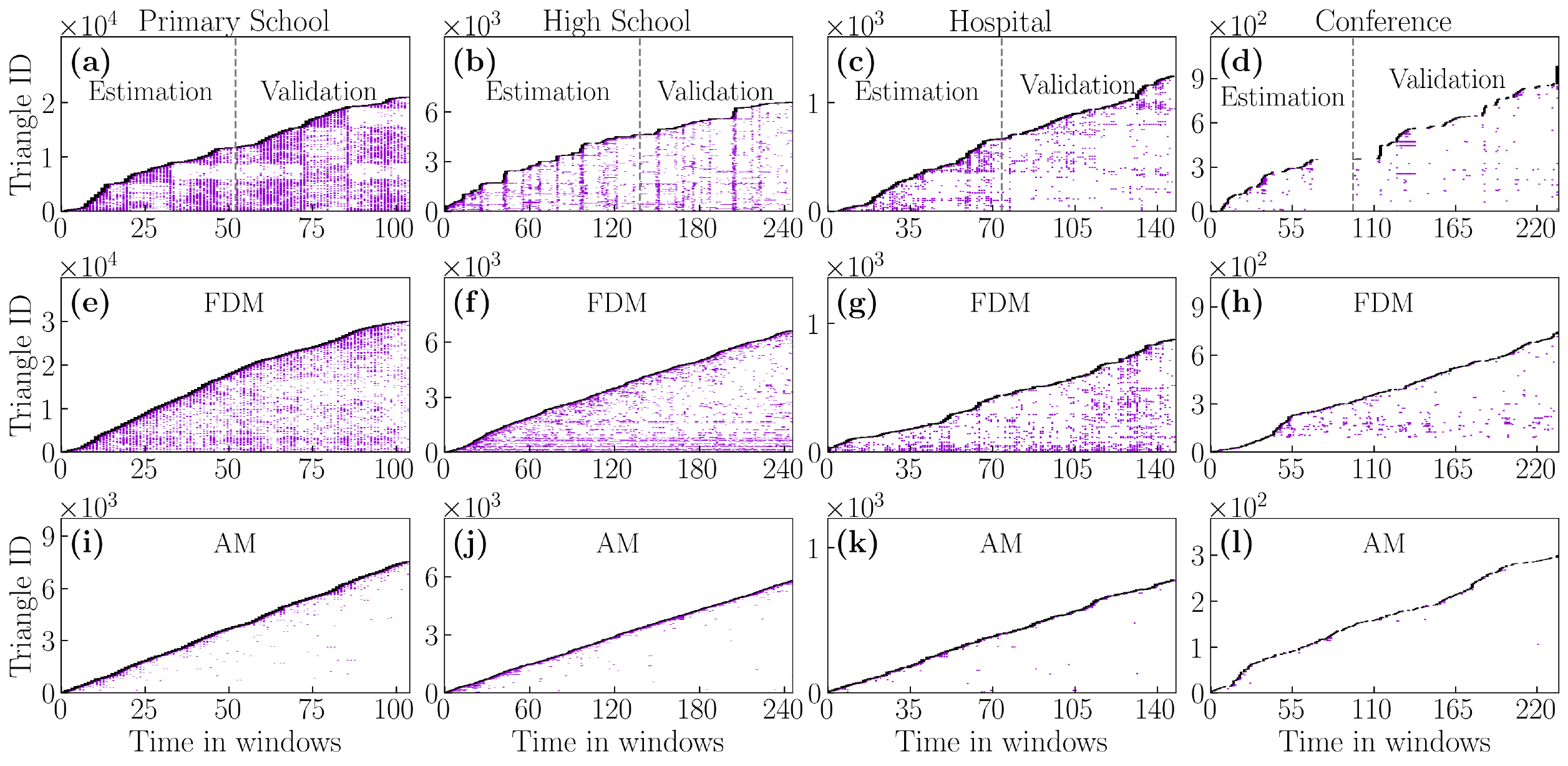}
	\caption{\textbf{Recurrence of groups in real-world and simulated networks.} (a-d) show the recurrent patterns of full triangles in different real-world networks. The gray dashed line separates the estimation and validation sections used in the model. (e-h) and (i-l) show the recurrent patterns of full triangles for the corresponding networks simulated by the FDM and AM models, respectively. In each figure,  the purple lines represent the recurrent full triangles, while the black ones correspond to the first occurrence of the triangles. We bin the $x$ axis into $10$ min intervals (30 time slots) as a time window and obtain an aggregated network snapshot for each time window. The $y$ axis shows the triangle IDs in each time window.}
	\label{figure1}
\end{figure*}
\begin{figure*}[!ht]
	\centering
	\includegraphics[width=1.0\textwidth]{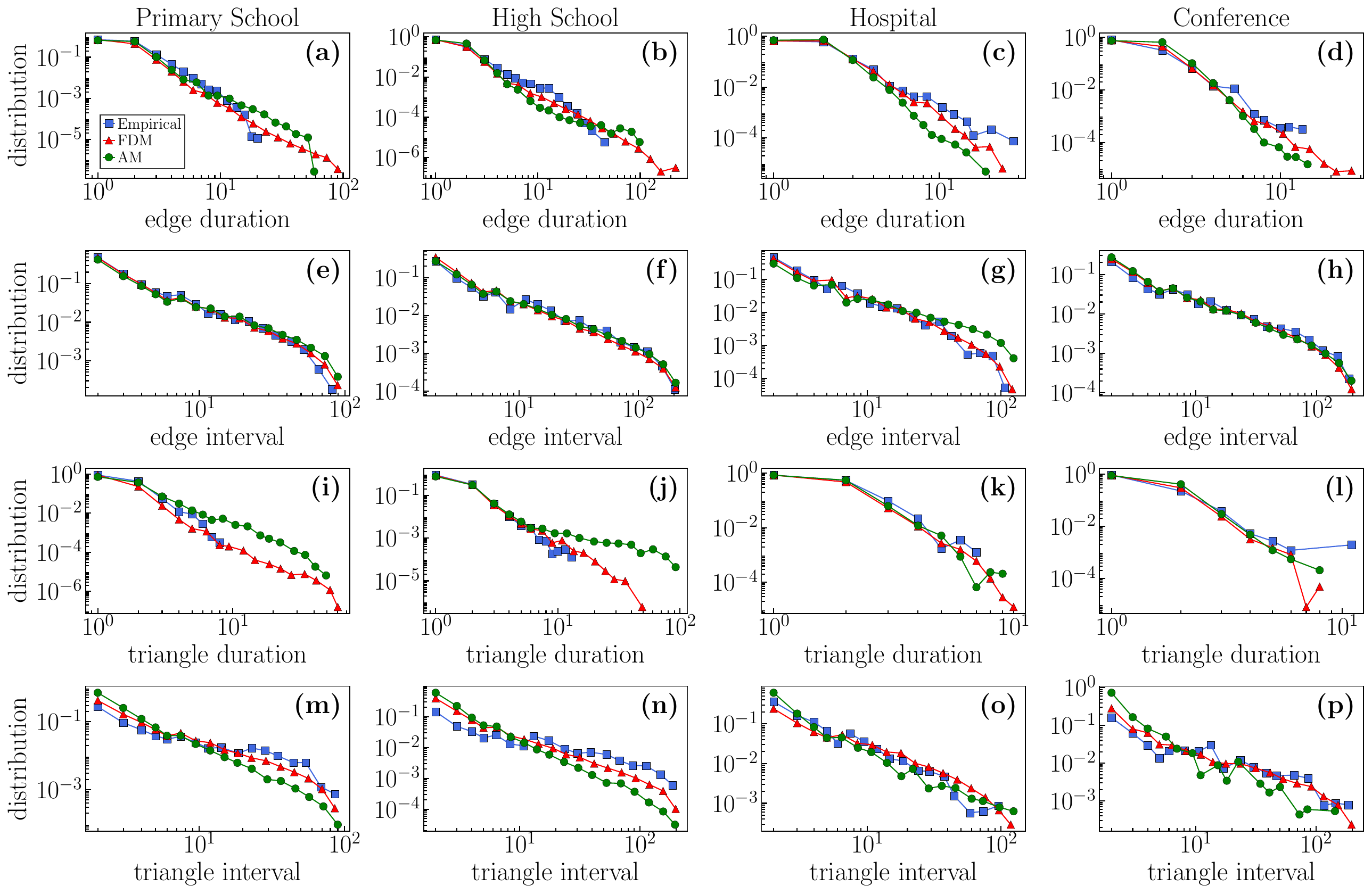}
	\caption{\textbf{Network properties of the real-world datasets and corresponding simulated networks.} Each column corresponds to a real network. (a-d) show the distribution of contact duration between a pair of nodes. (e-h) show the distribution of interval time between consecutive contacts of edges. (i-l) show the distribution of time duration between full triangles. (m-p) show the distribution of interval time between consecutive full triangles. All the results are measured based on the series of aggregated network snapshots. The simulated results of the models are averages over $20$ realizations.}
	\label{figure2}
\end{figure*}

\section{Model description}\label{Model}
\subsection{Force-directed motion (FDM) model }\label{FDM}

The FDM model was initially introduced to elucidate several key characteristics of face-to-face interaction networks, including the distributions of communication durations, weight distributions, recurrent component patterns, etc.~\cite{flores2018similarity}.
In the FDM model, individuals can move and interact with others in a two-dimensional Euclidean space (a square of size $L \times L$). Each individual's movement is not entirely random but is influenced by pairwise similarity forces. The individuals are assumed to inhabit a hidden similarity space, in which the distances between individuals act as similarity forces, guiding their movements within physical space and determining the interacting time.

In the hidden similarity space, $N$ individuals are positioned along a one-dimensional circle with radius 
$R = N / 2\pi$. An angular coordinate $\theta_i$ is randomly pre-assigned at interval $[0, 2\pi]$ for each individual $i$. The similarity distance is defined as 
$ s_{ij} = R \Delta \theta_{ij} $ for any two individuals 
$i$ and $j$, in which $ \Delta \theta_{ij} = \pi - \left| \pi - \left| \theta_i - \theta_j \right| \right| $ represents the angular distance between individuals 
$i$ and $j$. 

Time is discretized into slots in the model, and at the onset of each slot, individuals can have inactive or interacting states. When an inactive individual $i$ evolves active with a pre-assigned rate $r_i$, s/he can move within the slot. While interacting individuals are restricted to move unless they exit current interactions. On the other hand, each interacting individual $i$ has a probability $ P_i^e(t) $ of exiting their interaction at time $t$, where $ P_i^e(t) $ is expressed as:   

\begin{equation}\label{eq:1}
	P_i^e(t) = 1 - \frac{1}{\left| \Lambda_i(t) \right|} \sum_{j \in \Lambda_i(t)} e^{-s_{ij} / \mu_1},
\end{equation} 
where $\Lambda_i(t)$ represents the individuals' set that $i$ is now interacting with. $\mu_1$ stands a decay constant that regulates the average interaction duration.

At each time slot $t$, the position $( X_i^t, Y_i^t )$ of each moving individual $i$ is updated by  
\begin{equation}\label{eq:2}
	X_i^{t+1} = X_i^t + \sum_{j \in \Omega(t)} F_{ij} \frac{\left( X_j^t - X_i^t \right)}{\sqrt{\left( X_j^t - X_i^t \right)^2 + \left( Y_j^t - Y_i^t \right)^2}} + \varPsi_i^x,
\end{equation} 
\begin{equation}\label{eq:3}
	Y_i^{t+1} = Y_i^t + \sum_{j \in \Omega(t)} F_{ij} \frac{\left( Y_j^t - Y_i^t \right)}{\sqrt{\left( X_j^t - X_i^t \right)^2 + \left( Y_j^t - Y_i^t \right)^2}} + \varPsi_i^y,
\end{equation} 
where $\Omega(t)$ represents all moving and interacting individuals' set at the current time slot, and $F_{ij}$ indicates the degree of the attractive force between individuals $i$ and $j$. Note that $F_{ij}$ is a function of the similarity distance and is defined as:  
\begin{equation}\label{eq:4}
	F_{ij} = F_0 e^{-s_{ij}/\mu_2},
\end{equation}
where $F_0$ represent the magnitude of the force at $s_{ij} = 0$. The decay constant $ \mu_2>0 $ governs the influence of the force strength as $s_{ij}$ grows. In the motion equations, the displacement components of individual $i$ are given by $\varPsi_i^x = \nu \cos \alpha_i $ and $\varPsi_i^y = \nu \sin \alpha_i $, where $ \alpha_i $ is an angle sampled from the interval $[0, 2\pi)$ at random, and $ \nu \geq 0 $ denotes the importance of the arbitrary displacement. When $\nu = 0$, the motion becomes completely deterministic, while setting $ F_0 = 0 $ results in motion that behaves as a pure random walk. After an individual updates its position, it transitions to the interacting state if its distance to any other active individual is less than the interaction range $ d = 1 $; otherwise, it switches to the inactive state.

We give the procedures to tune the FDM parameters in Sec.~I in Supplementary Information (SI). 
We report the final parameters used in this work in Table~\ref{table1}. Note that multiple parameter sets may meet the requirements for each data set in the end. We show our results with one of the combinations as in Table~\ref{table1}. The results are robust if we choose another set of parameters after tuning (see Sec.~I and Figs.~S1 and S2 in SI).

\renewcommand{\thetable}{\arabic{table}} 
\newcommand{\thd}[1]{\multicolumn{1}{l}{#1}}
\begin{table*}[htbp] 
	\centering
	\caption{FDM model parameters and statistics for considered real-world networks. $N$ is the number of nodes to simulate. The parameter $T$ denotes the time slots to be simulated, corresponding to the time slots in the estimation section (i.e., observation part). $L$ is the square's side length that defines the boundaries of the two-dimensional Euclidean space. $\mu_{1}$, $F_{0}$, and $\mu_{2}$ are the FDM parameters used for simulating each empirical network. $\overline{n}$ and $\overline{l}$ are the average numbers of interacting individuals and edges per slot.} 
	\begin{tabular}{*{9}{l}}
		\toprule
		\thd{Data set}   & \thd{$N$} & \thd{$T$} & \thd{$L$} & \thd{$\mu_{1}$} & \thd{$F_{0}$} & \thd{$\mu_{2}$} & \thd{$\overline{n}$} & \thd{$\overline{l}$}\\
		\midrule
		Primary School          & 242                    &1555                           & 62                      & 0.85                  & 0.13                      & 0.82                  & 54.79                      & 38.99                                         \\			
		High School             & 327                    &4137                               & 94                      & 2.10                  & 0.40                      & 0.15                  & 45.84                      & 28.11                        \\
		Hospital                & 70                      &2200                            & 128                     & 0.68                  & 0.10                      & 1.12                  & 6.96                      & 4.63                   \\
		Conference              & 113                     &2874                             & 177                     & 2.10                  & 0.04                      & 1.27                  & 3.96                      & 2.41                     \\
		\bottomrule
	\end{tabular}
	\label{table1}
\end{table*}

\subsection{Attractiveness Model (AM)}\label{ATTR}

For comparison purposes, we also re-introduce the AM model here and examine whether it can reproduce the recurrent group patterns. In the AM model, each individual carries an activation probability $r_{i}$ and a value for the attraction $a_{i}$~\cite{starnini2013modeling,starnini2016model}. They are uniformly sampled in the range $[0,1]$. Initially, all individuals reside in a two-dimensional Euclidean space (a square of size $ L \times L $), where each individual can be in one of two states: active or inactive. An active individual moves within the space and engages in interactions with others, while an inactive individual neither moves nor participates in interactions.  

At each discrete time slot, an individual $i$ changes its state from inactive to active with rate $r_i$, while an active but isolated individual $j$ (i.e., not interacting with any other individual) evolves inactive with rate $1 - r_j$. Active individuals randomly walk in the space, moving in a random direction at a constant speed (displacement) during each time slot. When an individual encounters another individual within a distance $d=1$, they stop moving and begin to interact.   

The activation probability $r_{i} $ represents each individual's activity level in social events. In contrast, the global attraction value of an individual defines its ability to attract the interest of others and the probability of escaping from an interaction. For example, an individual $i $ that has stopped moving and is interacting with other individuals within a distance $d$ can continue moving with a rate of $1 - \max_{j \in \mathcal{S}_{i}} \{ a_{j} \}$, in which $\mathcal{S}_{i}$ is the collection of people interacting with $i$. In this sense, when interactions involve individuals with higher global attraction $a_{j} $, the interaction time tends to be longer.

We provide the procedures to tune the AM parameters in Sec.~I in SI. 
The model just needs to adjust the size of the Euclidean space $L$ (see Sec.~I in SI for more details). The values of $L$ using in AM in this work for the primary school, high school, hospital, and conference data sets are $48$, $76$, $38$, and $78 $, respectively.

\section{higher-order SIS spreading model and theoretical analysis} \label{SIS}
\subsection{Higher-order SIS spreading model}
To investigate how the recurrence of groups affects the spreading behaviors, we perform the higher-order SIS spreading  model~\cite{iacopini2019simplicial,wang2024correlation, matamalas2020abrupt} to real and synthetic temporal networks. 
To the best of our knowledge, the higher-order SIS spreading model was first proposed in Ref.~\cite{iacopini2019simplicial}, and this model is suitable and reasonable to simulate the information propagation on temporal networks. 

In the SIS propagation model, every node can occupy a state, Susceptible (S) or Infected (I)~\cite{keeling2008modeling}. State S indicates that the node is not receptive to the information or taking any action, while state I represents the node has accepted the information and is spreading it to its neighboring nodes. We perform the higher-order SIS spreading model on real and synthetic aggregated network snapshots as follows:

(i) We randomly select a fraction of nodes, $\rho_0$, to I states at the outset and set the remaining as S states.

(ii) The propagation process consists of pairwise and higher-order interactions at each step. For each time $t$, a node $i$ in the S state 
can become infected through classical pairwise interactions  with probability $\beta$ by its infected neighbors in the snapshot. In addition, if node $i$ is part of a triangle $[i, j, k]$ in the snapshot, and both of its neighbors $j$ and $k$ stay in I states, it may become infected through a higher-order interaction (the full triangle) with possibility $\beta_{\triangle}$. Note that when $\beta_{\triangle} = 0$, the model simplifies to the classical SIS model.

(iii) Each node in the state I will independently turn to the state S with probability $\mu$, regardless of whether the infected node interacted in the snapshot.

\subsection{A theoretical analysis with MMCA on temporal networks}\label{IV}

We use the MMCA to temporal networks and analyze our simulation results as described in Ref.~\cite{matamalas2020abrupt}. Let 
$p_i(t)$ represent the probability that node 
$i$ stays in the state I at time $t$. We then can express the corresponding infected density as:
\begin{equation}
	\rho(t)=\frac{1}{N}\sum\limits_{i=1}^Np_i(t). \label{eq:0}
\end{equation}
At time $t+1$, one can obtain the probability that a node $i$ gets infected as:
\begin{equation}\label{eq:5}
	p_ {i}(t+1)=(1- g_i  (t)g_{i,  \triangle } (t))(1-  p_ {i}  (t))+(1-  \mu  )  p_ {i}  (t) ,
\end{equation}
where $g_i(t)$ represents the probability that node $i$ does not become infected from a link with its neighbors in the snapshot (i.e., pairwise interaction):
\begin{equation}\label{eq:6}
	g_i (t)=\prod_{j\in \Lambda_{i}\left ( t \right )   }^{} (1- \beta p_ {j} (t)) ,
\end{equation}
where $\Lambda_i(t)$ represents the neighbors' set of node $i$ at time $t$. Besides, $g_{i,\triangle}(t)$ represents the probability that node $i$ does not get infected through interactions in its 2-simplices in the snapshot:
\begin{equation}\label{eq:7}
	g_{i,  \triangle } \left ( t \right ) = \prod_{j,l\in\Theta_{i}\left ( t \right )   }^{} (1-  \beta _ {\triangle }   p_ {j}  (t)p_ {l}  (t)),
\end{equation}
where $\Theta_i(t)$ represents the sets of 2-simplices including node $i$ at time $t$. It is important to note that $\Lambda_i(t)$ and $\Theta_i(t)$ are functions of time~\cite{chowdhary2021simplicial}, which enables the extension of the MMCA method to temporal networks.

\section{Result}\label{results}
\subsection{The recurrent patterns of groups in empirical data}\label{Reappearance}
\begin{figure*}[!ht]
	\centering
	\includegraphics[width=1.0\textwidth]{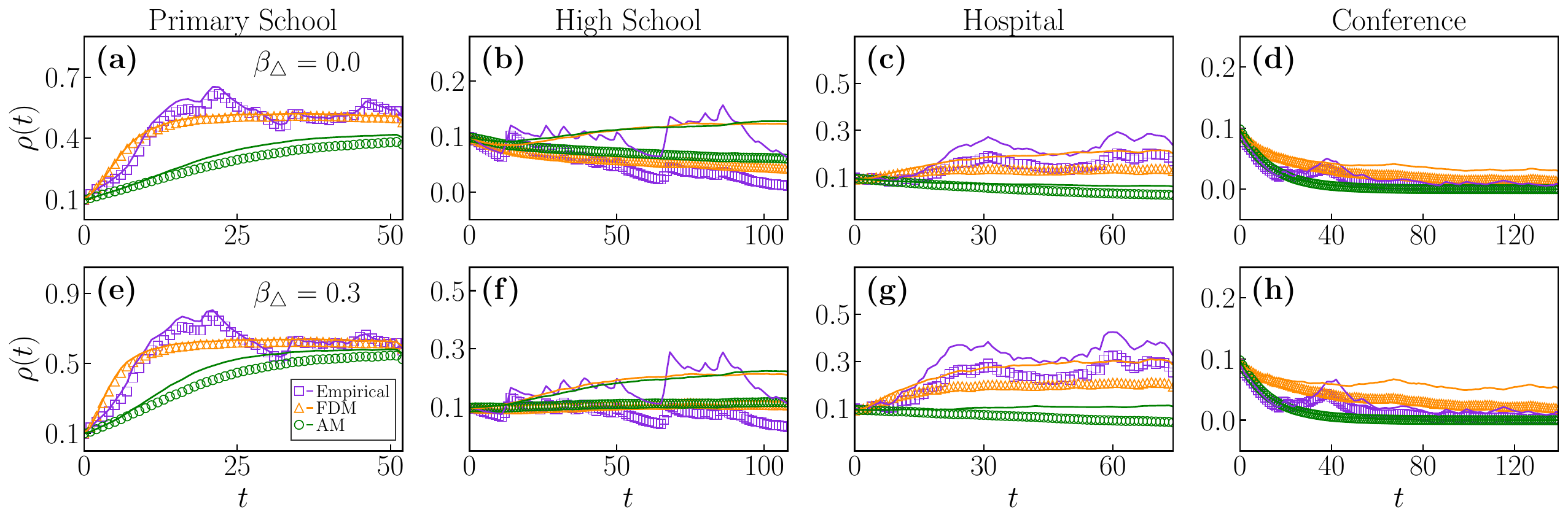}
	\caption{\textbf{The time evolution of infected nodes' densities in real and simulated networks.} Each column represents a real network. The upper and lower rows correspond to the cases without higher-order interactions (i.e., $\beta_{\bigtriangleup} = 0.0$) and with higher-order interactions (i.e., $\beta_{\bigtriangleup} = 0.3$), respectively. Symbols and solid lines illustrate the outcomes of the simulations and the MMCA, respectively. All results were derived from 10,000 independent realizations comprising 20 distinct network configurations and 500 initial conditions for each configuration. The parameters are $\mu = 0.1$, $\beta = 0.1$, and $\rho_0 = 0.1$.}
	\label{figure3}
\end{figure*}

To better observe the formation of groups in empirical data, 
we bin the empirical data nonoverlapping with $10$ min interval ($30$ time slots per bin) as a time window (i.e., an observation interval) and aggregate the networks in each bin to a snapshot. This way, we obtain a series of aggregated network snapshots for each data set. For simplicity, we only consider a group constituted with full triangles among three nodes. The results are robust for considering groups formed with more nodes, i.e., triangles, tetrahedrons, pentachorons, etc. (See Fig.~S3 in SI for more details). For an aggregated network snapshot in a time window, we 
extract all the full triangles and assign an identity (ID) to the first occurrence of full triangle. The ID is the same in the subsequent network snapshots if a triangular group is constituted by the same nodes $[i,j,k]$. 

In Figs.~\ref{figure1}(a-d), we find the recurrent patterns of groups in the real-world networks. In each pattern, a black line represents the first occurrence of a triangle, while the purple lines depict recurrent triangles, i.e., the groups that have occurred at least once. The recurrent patterns of full triangles are also held if we aggregate the networks with $5$ or $30$ min interval (see Fig.~S4 in SI). Note that recurrent patterns also exist if we consider the groups constituted with different numbers of nodes, such as triangles, tetrahedrons, pentachorons, etc. (See Fig.~S3 in SI). These results indicate that the recurrence of groups is a typical phenomenon in human face-to-face interaction networks. 

\subsection{The FDM model reproduces the recurrent group patterns}

The recurrent patterns of full triangles observed in the real temporal networks excite our great interest and inspire us to explore the underlying mechanisms. We here extend the 
FDM and AM models and try to reproduce the recurrent group patterns. We separate the real data with estimation and validation sections as demonstrated by the gray dashed lines in Figs.~\ref{figure1}(a-d). We use the estimation section to adjust the model's parameters (see the model parameters tuning in Sec.~I in SI) and validate them in the validation section. Note that we also bin the simulated data nonoverlapping with $10$ min interval as a time window (i.e., $30$ time slots in each bin) and aggregate the networks in each bin to a snapshot. From Figs.~\ref{figure1}(e-h), we can easily observe that the FDM model can replicate similar patterns of recurrent groups. However, the AM could not capture these patterns very well, in which only rarely triangles occur as recurrent phenomena.

Besides the recurrent patterns of groups, we also investigate the temporal network properties of the real data and corresponding simulated networks. 
In Figs.~\ref{figure2}(a-h), we show the distributions of contact duration between a pair of nodes and the interval time between consecutive contacts of edges. We find that both FDM and AM models can effectively replicate these distributions as real data. In addition, we compare the real with synthetic temporal networks about the distributions of time duration between full triangles and the interval time between consecutive full triangles in Figs.~\ref{figure2}(i-p). We can see that the FDM model can capture the properties of time duration between triangles and the interval time between consecutive triangles better than AM in temporal networks. 
Furthermore, the FDM model also effectively captures a variety of additional topological and structural properties of the real temporal networks (see the distributions of edge weights and node strengths in Fig.~S5 in SI). 

These results imply that the forces arising from similarity distances in the hidden spaces in the FDM model offer a compelling framework for explaining the recurrent patterns of groups observed in human interaction networks. These forces govern the movement of individuals within the physical space and modulate the time of their interactions. Therefore, the FDM model is capable of capturing a broad spectrum of many fundamental characteristics of human face-to-face interaction networks.
\subsection{The FDM model predicts the information spreading behaviors} 

\begin{figure*}[!ht]
	\centering
	\includegraphics[width=1.0\textwidth]{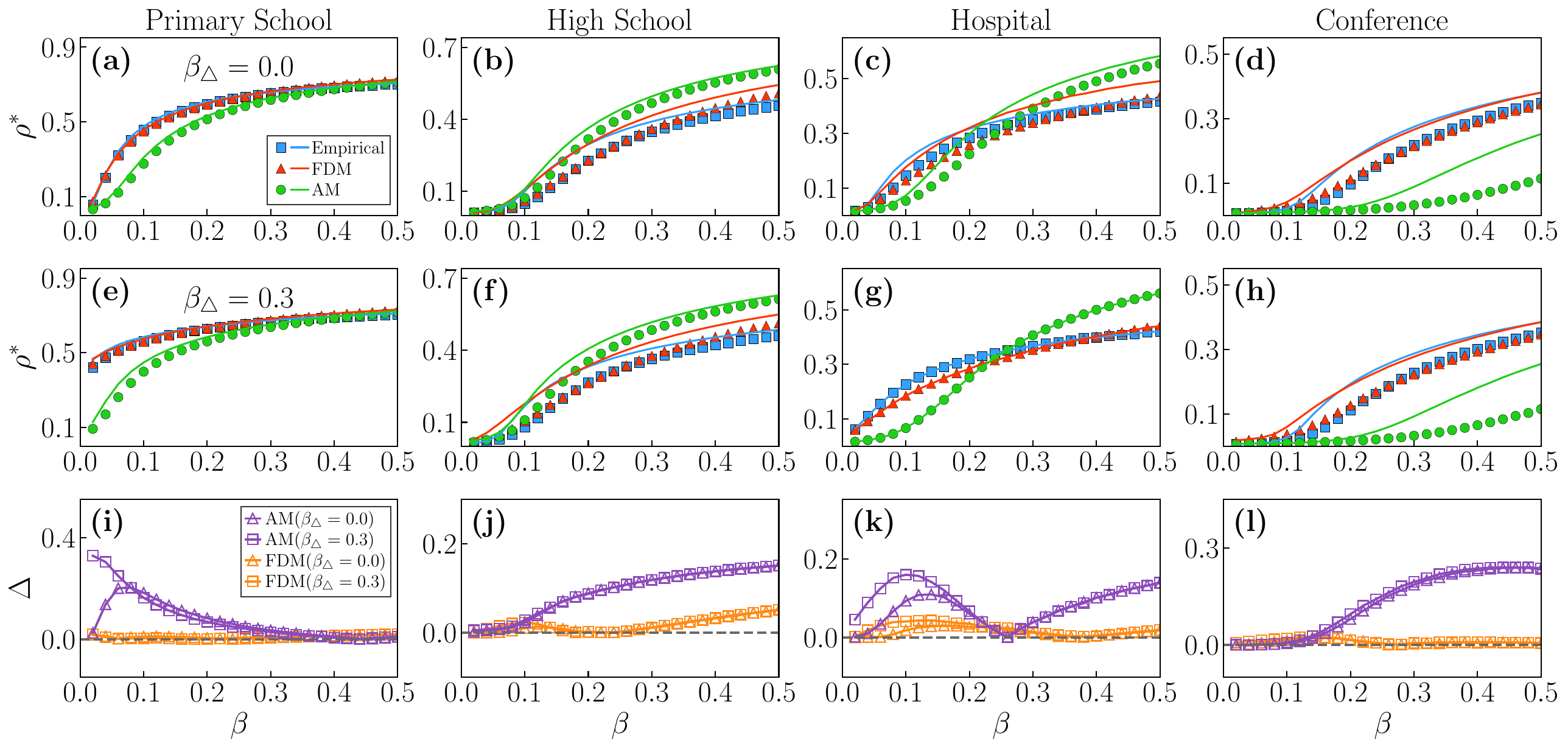}
	\caption{\textbf{The final spreading range $\rho^*$ as a function of infection probability $\beta$.} Each column stands for a real network. The first and second rows represent the results without and with higher-order interactions (i.e., $\beta_{\bigtriangleup} = 0.0$ and $\beta_{\bigtriangleup} = 0.3$), respectively. The third row shows the absolute errors of the final spreading range $\rho^*$ between real networks and models. The symbols and lines in (a-h) represent the model simulation and MMCA results, respectively. Other parameters are same as Fig.~\ref{figure3}. }
	\label{figure4}
\end{figure*}
We subsequently investigate whether the information spreading processes in real and simulated networks exhibit similar behaviors. To characterize the spreading behavior quantitatively, we introduce the macroscopic order parameter $\rho(t)$, representing the density of infected individuals at time $t$. Additionally, we employ $\rho^*$ to denote the information coverage of the spreading process in the end (i.e., the final spreading range), obtained by averaging the infected node density over all snapshots.
We perform the information spreading model in the validation section of human face-to-face interaction data. At the same time, we use the synthetic networks from the FDM and AM models to predict the spreading behaviors.

Figure~\ref{figure3} shows the time evolution of infected nodes' densities $\rho(t)$ in real and simulated networks with a pairwise infection probability of $\beta = 0.1$. 
It is evident that irrespective of the incorporation of higher-order interactions (i.e., $\beta_\Delta=0$ or $\beta_\Delta=0.3$), the propagation results obtained from the FDM model closely resemble those of the real network. In contrast, the propagation results from the simulated networks based on the AM exhibit significant discrepancies compared to the real ones. The theoretical analysis by the MMCA confirms all the numerical simulations.

To observe the information spreading behaviors more systematically, we show the final spreading coverage $\rho^*$ versus infection probability $\beta$ in Figs.~\ref{figure4}. Unsurprisingly, the spreading coverage is similar in real data and the FDM model, as the structures generated by the FDM model are comparable. In contrast, the spreading results in AM show a discrepancy from the real data to some extent. Furthermore, we find that when considering the higher-order interactions $\beta_{\bigtriangleup} = 0.3$, the difference between AM and real networks tends to exacerbate with small $\beta$ (see the absolute errors between the final infected fraction $\rho^*$ in real networks and models in Figs.~\ref{figure4}(i-l)). 
The theoretical analysis performed by the MMCA fully corroborates the results obtained from the numerical simulations. These results indicate that the FDM model can predict the information spreading behaviors in human face-to-face interaction networks.
\begin{figure*}[t]
	\centering
	\includegraphics[width=1.0\textwidth]{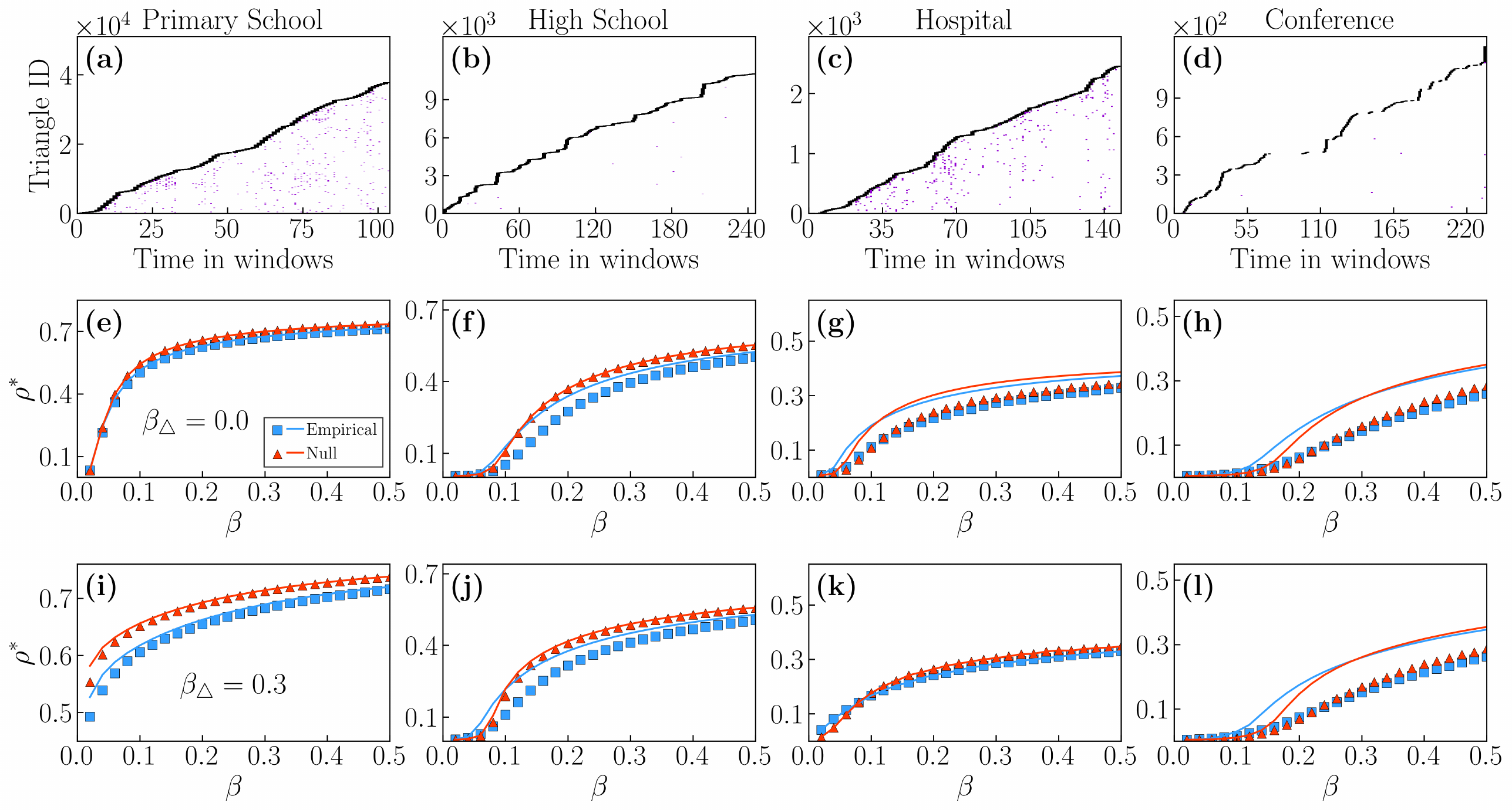}
	\caption{\textbf{Recurrence of full triangles suppresses the information spreading.}	Each column stands for a real network. (a-d) show the null model's recurrent patterns of full triangles. In each figure,  the purple lines represent the recurrent full triangles, while the black ones correspond to the first occurrence of the triangles. (e-l) show the $\rho^*$ as a function of $\beta$ for each real network. The second and third rows represent the results without and with higher-order interactions (i.e., $\beta_{\bigtriangleup} = 0.0$ and $\beta_{\bigtriangleup} = 0.3$), respectively. The symbols and lines illustrate the model simulation and MMCA results, respectively. Other parameters are same as Fig.~\ref{figure3}. 
	}
	\label{figure5}
\end{figure*}

\subsection{The influence of recurrent groups on information spreading} 

The above findings indicate that the recurrence of groups significantly impacts information propagation. An interesting question arises: Does the recurrence of full triangles speed up or suppress the information propagation? We can not get the answer by comparing the spreading results between the FDM and AM models, as the number of triangles is not the same in each snapshot in the FDM and AM models.  

To overcome it, we construct a null model on real temporal networks so that we have the same number of edges and triangles in each snapshot. Specifically, we swap the nodes' labels randomly in each snapshot (see the construction of the null model in Sec.~II in SI for more details). By doing so, we preserve the same number of groups and edges in each snapshot but decrease the recurrence of groups. Figures~\ref{figure5}(a-d) show the null model's recurrent patterns of full triangles. Similar to Fig.~\ref{figure1}, black lines represent newly appeared triangles, and purple lines denote recurrent ones. Comparing with Figs.~\ref{figure1}(a-d), we can see that the null model has significantly reduced the recurrence of full triangles (see the decreasing of purple lines and the increasing of the black lines in the figures and note the $y$-axis scale) while maintaining the number of the edges and groups consistent with the real networks. 
Next, we perform the higher-order SIS spreading model on the  snapshots from the real data and their corresponding null models. We show the numerical simulations and the corresponding MMCA theoretical analysis in Figs.~\ref{figure5}(e-l). We observe consistent results across all data sets ---  compared to the null model, real-world networks with more group recurrence exhibit smaller information spreading range. Moreover, this phenomenon is more remarkable at higher-order interactions (i.e., $\beta_{\bigtriangleup} = 0.3$). In other words, the recurrence of groups tends to suppress information spread in real-world face-to-face interaction networks. A plausible explanation is that the information tends to remain localized within a small cohort of individuals who engage in frequent interactions, thus hindering its diffusion to other individuals outside this group. 

\begin{figure}[!ht]
	\centering
	\includegraphics[width=0.48\textwidth]{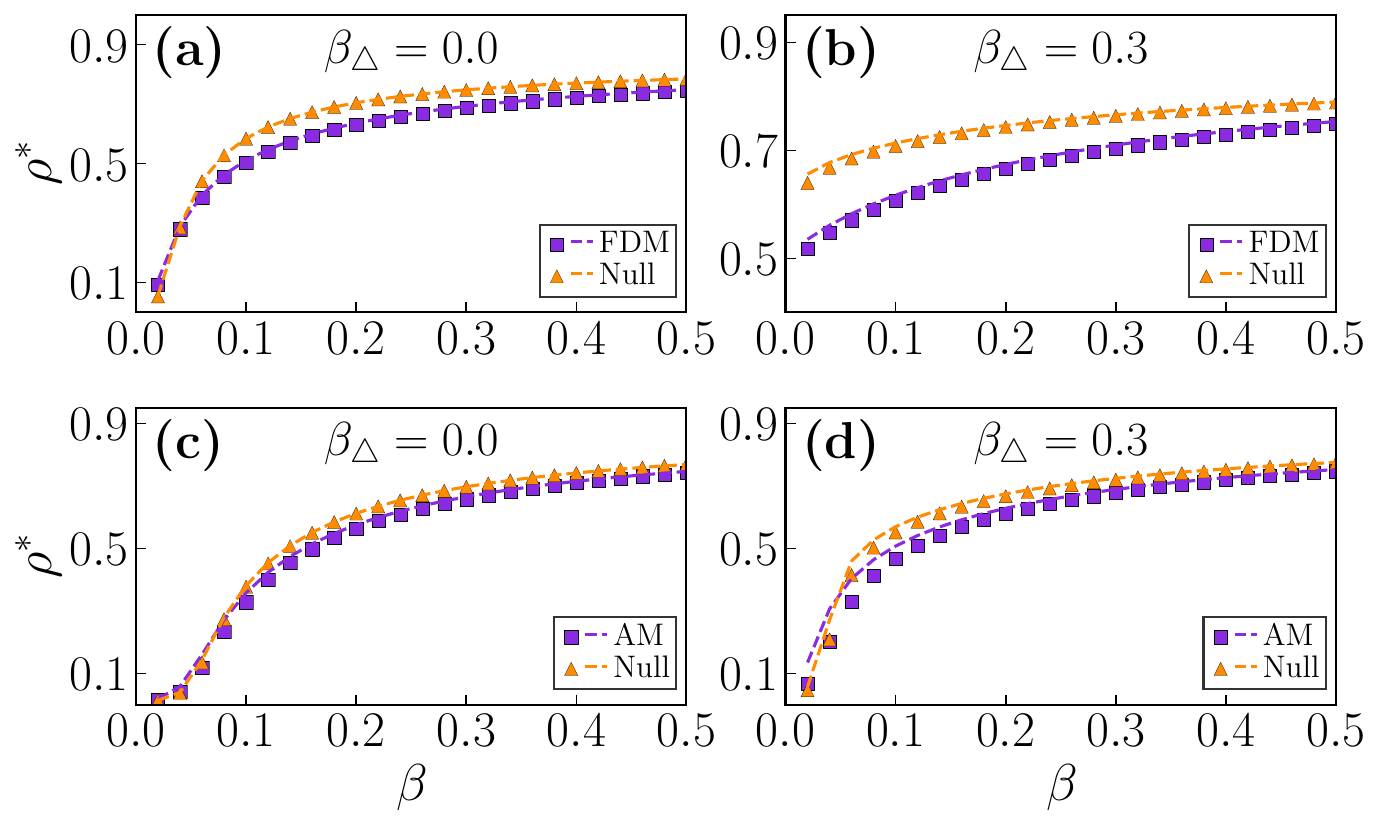}
	\caption{\textbf{Recurrence of full triangles inhibits the information spreading in the FDM and AM models.} 
		Each figure shows the $\rho^*$ as a function of $\beta$ for simulated networks by the FDM or AM model and their corresponding null models. The simulated networks are generated using the parameters in primary school. The first and second columns show the spreading range $\rho^*$ without higher-order interactions (i.e., $\beta_{\triangle} = 0.0$) and with higher-order interactions (i.e., $\beta_{\triangle} = 0.3$), respectively. Symbols and dashed lines in the figure represent the simulation and MMCA results, respectively. Other parameters are same as Fig.~\ref{figure3}. }
	\label{figure6}
\end{figure}

Besides the network snapshots from the real data, we also perform the simulations to verify the above conclusion based on the FDM and AM models.
We simulate $20$ networks with the FDM and AM models using the parameters in primary school (see the robust results with the parameters from the other data sets in Figs.~S6-S8 in SI). We generate the corresponding null model for each simulated network and perform the information spreading dynamics on top of them. As shown in Figs.~\ref{figure6}(a) and (b),  we observe that for the FDM model, where group recurrences are more frequent, the propagation results of the simulated networks and their null models show very significant differences, which is similar to the observation in real data that the recurrent groups reduce the information spreading. In contrast, although the number of recurrent triangles is rare in AM, the spreading range is still slightly extensive in its null model in Figs.~\ref{figure6}(c) and (d). It's just that the differences in propagation behaviors between the AM and its null model aren't that apparent. In short, we find that for the networks simulated by the FDM and AM models, the previously mentioned conclusion still holds---the recurrence of groups will inhibit the information spreading in time-varying networks, and the higher-order interactions will make this phenomenon more pronounced. 

\section{Conclusion}\label{Conclusion}
Face-to-face interaction networks represent the underlying structure of social interactions within human gatherings and serve as the substrate for processes such as epidemic propagation and the dissemination of information or gossip.
However, the mechanisms underlying group formation in face-to-face interaction networks remain inadequately understood. Additionally, the impact of higher-order interactions, which arise from group dynamics in temporal networks, on the information-spreading process has yet to be thoroughly investigated.

In this article, we analyzed four real-world human face-to-face interaction networks and discovered the phenomenon of recurrent triangular group formations. Moreover, we extended the FDM model to reproduce the recurrent patterns of groups and a wide range of other prominent properties. The forces arising from similarity distances within metric spaces in the FDM model offer a fundamental explanation for the observed recurrent triangular group dynamics in real networks. Furthermore, we showed that the FDM model could predict information-spreading behaviors under higher-order interactions. Finally, we demonstrated that the recurrent triangular groups significantly impact information spreading, especially under higher-order interactions. Notably, the recurrence of triangular groups will inhibit the spread of information in temporal networks, and the higher-order interactions will make this phenomenon more pronounced.


In conclusion, our study enhances the understanding of human behavior through the lens of group formation and interaction dynamics. The insights gained, and the proposed model provides valuable perspectives on the mechanisms underlying information propagation in the presence of higher-order interactions in real-world scenarios.

\section*{Acknowledgments}
This work was supported by National Natural
Science Foundation of China (Grants No.~12305043 and No.~12165016), the Natural Science Foundation of Jiangsu Province (Grant No.~BK20220511), 
M. Z. appreciates the support from the Jiangsu Specially-Appointed Professor Program.

\section*{CRediT authorship contribution statement}
\textbf{Liang Yuan:} Methodology, Software, Visualization, Writing-original draft. 
\textbf{Jiao Wu:} Conceptualization, Software, Writing-original draft, Funding acquisition.
\textbf{Kesheng Xu:} Writing-original draft, Funding acquisition. 
\textbf{Muhua Zheng:} Methodology, Software, Conceptualization, Visualization, Data Curation, Writing-original draft, Funding acquisition, Supervision

\section*{Declaration of competing interest}
The authors declare that they have no known competing financial interests or personal relationships that could have appeared to influence the work reported in this paper.

\section*{Data availability}
The code and data obtained for
this study are available via the Zenodo platform at  \href{https://doi.org/10.5281/zenodo.14799110}{\color{blue}{https://doi.org/10.5281/zenodo.14799110}}.


\end{document}


\title{Supplementary Information for\\ The recurrence of groups inhibits the information spreading under higher‑order interactions}
	\date{\today}
	\author{Liang Yuan}
	\affiliation{School of Physics and Electronic Engineering, Jiangsu University, Zhenjiang, Jiangsu, 212013, China}
	
	\author{Jiao Wu}
	\affiliation{School of Mathematical Sciences, Jiangsu University, Zhenjiang, Jiangsu, 212013, China}
	
	\author{Kesheng Xu}
	\affiliation{School of Physics and Electronic Engineering, Jiangsu University, Zhenjiang, Jiangsu, 212013, China}
	
	\author{Muhua Zheng}
	\email[]{zhengmuhua163@gmail.com}
	\affiliation{School of Physics and Electronic Engineering, Jiangsu University, Zhenjiang, Jiangsu, 212013, China}
\maketitle

\begin{spacing}{1.2}
\tableofcontents
\end{spacing}

\renewcommand\thefigure{S\arabic{figure}}
\renewcommand\thetable{S\arabic{table}}
\renewcommand\theequation{S\arabic{equation}}

%
%
%
\section{Model parameters}
\subsection{Parameter tuning for FDM model} 
There are six key parameters in FDM~\cite{flores2018similarity}: (i) $N$, the number of individuals to simulate; (ii) $T$, the number of time slots to simulate; (iii) $L$, which defines the area of the two-dimensional Euclidean space in which agents move and interact (an $L \times L$ square); (iv) $\mu_1$ (as defined in Eq. (1) of the main text), which governs the average duration of interactions between individuals;  and (v) and (vi) $F_0$ and $\mu_2$ (as defined in Eq. (4) of the main text), which control the expected displacement of individuals due to attraction forces, as well as the abundance and size of components in the system. Besides these parameters, a warmup period, denoted as 
$T_{\text{warmup}}$, is needed, which refers to the simulation phase during which the average number of interacting individuals per time step stabilizes. All network properties are measured after the warmup period. This phase is necessary to allow individuals that are initially distributed uniformly in the Euclidean space to move towards each other in the similarity space, thereby enabling those individuals that are spatially close in the similarity space to become closer in the Euclidean space. In the following, we discuss how these parameters are calibrated in the simulated counterparts of each real network, with the corresponding values provided in Table 1 in the main text. 

Parameter $N$ represents the number of simulated nodes, corresponding to the total number of individuals participating in the interactions in each data set. $T$ represents the number of time slots to simulate, corresponding to the number of time slots in estimation section (i.e., observation part). $T_{\text{warmup}}$ is fixed as long as the average number of interacting individuals per slot stabilizes~\cite{flores2018similarity}. We report it in Table.~S1 for each data set. At the same time, we fix $\nu=1$ and $d=1$.


For the adjustment of the parameters $L$, $\mu_{1}$, $F_{0}$, and $\mu_{2}$, we first follow the tuning methodology outlined in the supplementary materials of the Ref.~\cite{flores2018similarity} and generate plenty of synthetic networks to determine the approximate initial range of these parameters. Note that all the parameter tuning is based on the observation part of real-world networks. 
More specifically, we adjust $\mu_{1}$ to ensure that the average contact duration in the simulation is approximately the same as in the real dataset (as $\mu_{1}$ increases, the average contact duration becomes longer). The parameters $F_{0}$ and $\mu_{2}$ are tuned to match the average number of recurrent groups over 10-minute intervals in the simulation to the real dataset while also ensuring that the size of the largest group formed is similar to that in the dataset (As $ \mu_2 $ increases, larger connected components begin to form, eventually merging into a giant connected component. A similar phenomenon can be observed when $ F_0 $ increases, as the magnitude of deterministic motion becomes larger compared to random motion. In this case, if $ \mu_2 $ is not sufficiently small, it may also lead to the eventual merger into a giant connected component. To avoid merging into a giant connected component, when one of these two parameters increases, the other should decrease). The parameter $L$ is adjusted so that the average degree in the time-aggregated network aligns with the value observed in the real network (a larger $L$ leads to a more minor average degree). Following the above tuning methodology, a relatively narrow initial range for these parameters can be determined as in Table~S1.

\renewcommand{\arraystretch}{1.0} 
\newcommand*{\thd}[1]{\multicolumn{1}{l}{#1}}
\begin{table*}[!t] 
	\begin{ruledtabular}		
		\centering
		\label{tab:S1}	
		\caption{The initial parameter ranges used in FDM model. $L$ is the side length of the two-dimensional Euclidean space, $\mu_{1}$ and $\mu_{2}$ are two decay constants, and $F_{0}$ is the force magnitude when $s_{ij} = 0$, $T_{\text{warmup}}$ is the warmup period, $\overline{c}$ is the total number of recurrent groups.}
		\begin{tabular}{*{7}{l}}			
			\thd{Data set}   & \thd{$L$} & \thd{$\mu_{1}$} & \thd{$F_{0}$} & \thd{$\mu_{2}$} & \thd{$T_{\text{warmup}}$} & \thd{$\overline{c}$}\\
			\hline
			Primary School                           & [55,65]                    & [0.3,1.0]             & [0.10,0.25]                    & [0.20,1.20]             &    2000             & 681      \\			
			High School                               & [90,100]                    & [1.6,2.3]             & [0.25,0.55]                    & [0.01,0.80]            & 6500                  & 834      \\
			Hospital                                 & [110,130]                   & [0.6,1.0]             & [0.10,0.15]                    & [0.70,1.50]             &2500                 & 205       \\
			Conference                               & [90,190]                   & [2.0,3.0]           & [0.02,0.04]                     & [0.80,1.80]          & 6000                & 35           \\			
		\end{tabular}
	\end{ruledtabular}
\end{table*}

In addition, the parameter tuning process requires comparison with three statistical metrics in the real network: the average number of interaction nodes per time slot ($\overline{n}$), the average number of edges per time slot ($\overline{l}$), and the total number of recurring groups ($\overline{c}$). Note that to reduce time complexity, we do not split the groups and count $\overline{c}$ for the groups
with $k$-simplex and $k\in [2,\infty)$, i.e., $\overline{c}$ count the groups with triangles, tetrahedrons, pentachorons, etc. Specifically, every group only counts once within 10-minute intervals so that we can ignore short-term correlation and reduce time complexity. Then, $\overline{c}$ is calculated as the sum of the groups that appear more than once over all time windows.
With the initial range of $L$, $\mu_{1}$, $F_{0}$, and $\mu_{2}$ on hands, we perform the following tuning procedures:

\textbf{1. Generating synthetic temporal networks.} Using the initial ranges determined above, we generate a set of parameter lists by a small increment. Here we set the increment as $\Delta L=1$, $\Delta \mu_{1}=0.01$, $\Delta F_{0}=0.01$ and $\Delta \mu_{2}=0.01$. We would go through every value in the parameter lists and generate $20$ synthetic temporal networks for each combination of the parameters.
%
%

\textbf{2. Filtering Parameters:} For each combination of parameters, we firstly keep it if the $\overline{n}$ and $\overline{l}$ values of the simulated networks fall within an error margin of $0.2$ with the corresponding values of the real network. Next, we filter the parameters based on the average group count $\overline{c}$. We keep the parameter sets if the $\frac{|\overline{c}_{\text{simulated}}-\overline{c}_{\text{real}}|}{\overline{c}_{\text{real}}}<0.2$. Finally, we filter the parameter sets based on the distribution of the average contact duration and the distribution of maximum group size within a 10-minute interval in the simulated networks. We maintain the parameter sets with these two properties of the simulated networks closely match the real ones.

\textbf{3. Parameter Determination:} Based on the filtering method described above, multiple parameter sets may meet the requirements for each data set. The main text shows our results with a combination of parameters. We also find that the results are robust if we choose another set of 
parameters after the filtering (see Tables~S2 and S3 for the parameters and the results in Figs.~\ref{FDM1} and~\ref{FDM2}).

\subsection{Parameter tuning for AM} 
The model includes three parameters: $ N $, $ T $, and $ L $. The warmup phase is no longer required. Similarly, we directly fix the parameters $ N $ and $ L $. $ N $ represents the number of simulated nodes, corresponding to the total number of individuals participating in the interactions in each data set. $ T $ represents the number of time slots to simulate, corresponding to the number of observation time slots.

We obtain the model parameter $ L $ such that the errors of average $ \overline{n} $ and $ \overline{l} $ between the simulated and actual networks are within 0.2. In this way, the values of $ L $ for the primary school, high school, hospital, and conference data sets are $ L = 48, 76, 38, 78 $, respectively. 


\section{Construction of the Null model}
The null model must ensure that the generated network maintains the same number of nodes and edges in each window as the original network. Additionally, the number of groups of different sizes within each window must remain consistent. Here, we first count and store the indices of interacting nodes (i.e., not isolated nodes) in each window of the original network into a list. Then, we shuffle the index of nodes in this list and store them in a new list. Finally, by replacing the index of nodes in the list in the original networks, we obtain the networks by the null model. In short, we swap the interacting nodes' labels randomly in each snapshot.

\section{Supplementary figures}
\begin{figure*}[t]
	\centering
	\includegraphics[width=1.0\textwidth]{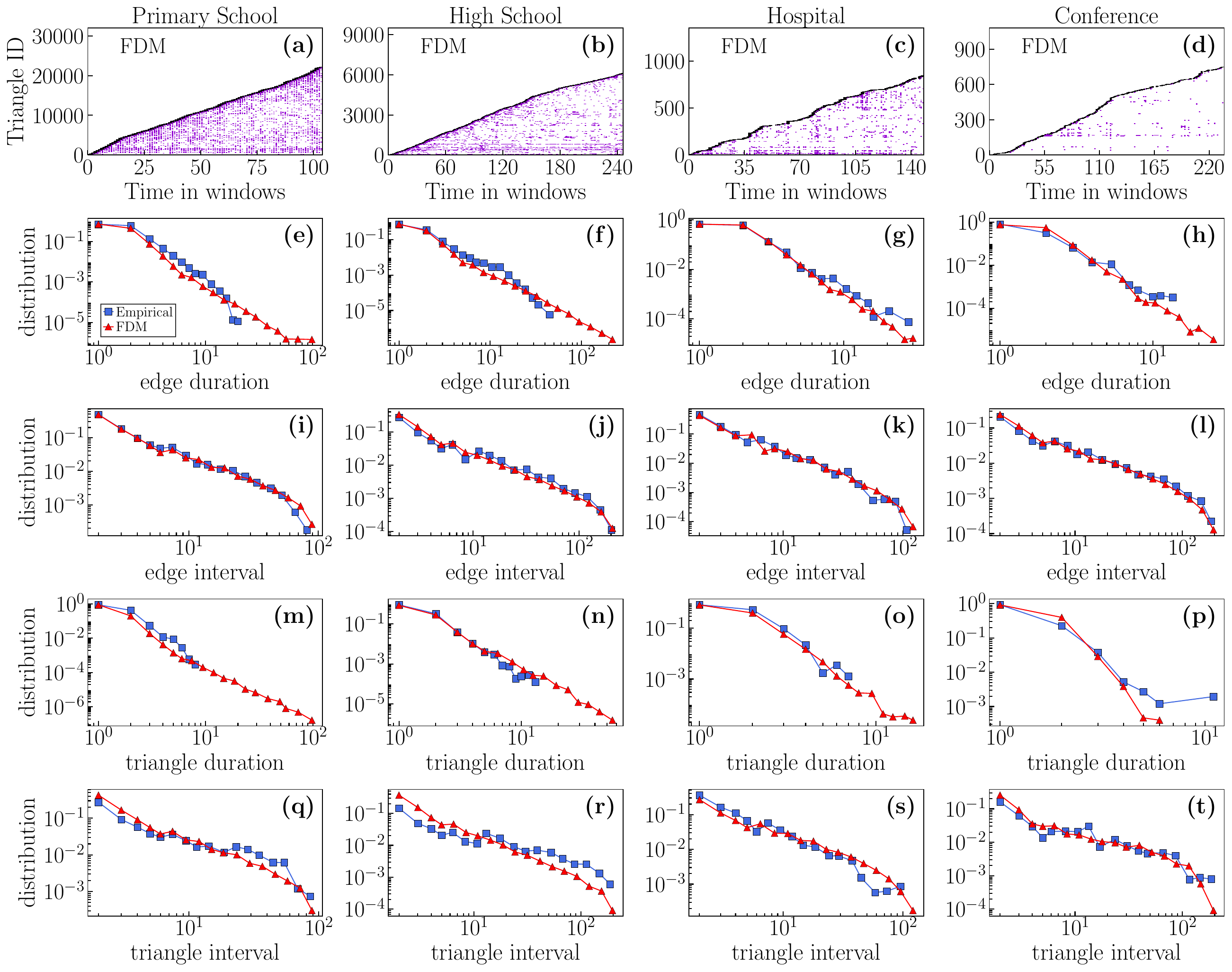}
	\caption{\textbf{The FDM model with different parameters still can capture the observed recurrent patterns of groups and many crucial features of real data.}  (a-d) show the recurrent patterns of full triangles from the FDM model with parameters in Table.~S2. (e-t) show the corresponding network properties as Fig.~2 in main text.}
	\label{FDM1}
\end{figure*}

\begin{figure*}[t]
	\centering
	\includegraphics[width=1.0\textwidth]{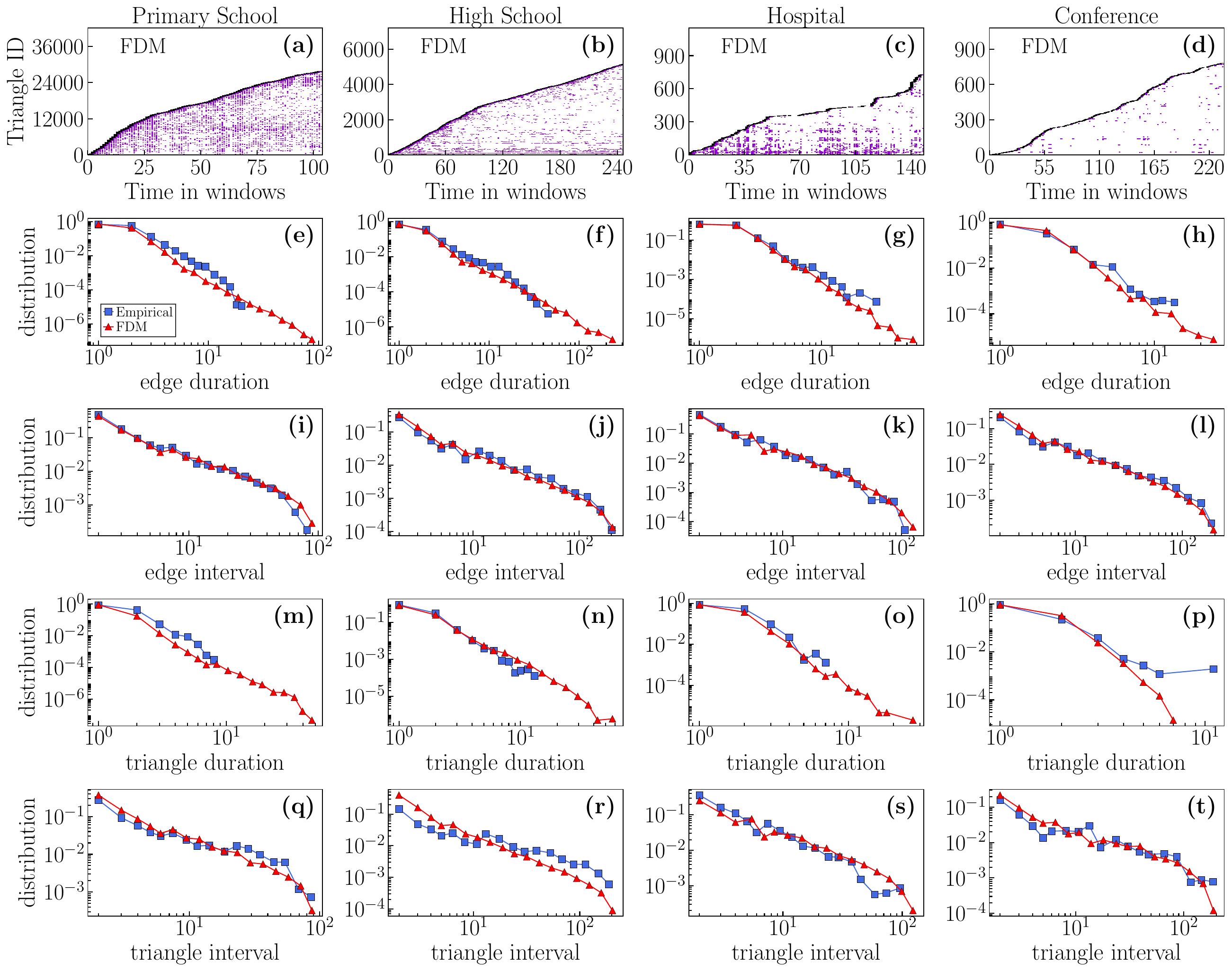}
	\caption{\textbf{The FDM model with different parameters still can capture the observed recurrent patterns of groups and many crucial features of real data.}  (a-d) show the recurrent patterns of full triangles from the FDM model with parameters in Table.~S3. (e-t) show the corresponding network properties as Fig.~2 in main text.}
	\label{FDM2}
\end{figure*}

In this section, we present additional figures to provide further evidence supporting our findings. 
We show that the results from the FDM model are robust if we choose another set of parameters after the filtering. That is, the FDM model with different parameters can still capture the observed recurrent patterns of groups and many crucial features of real data. See Tables~S2 and S3 for the parameters and the results in Figs.~\ref{FDM1} and~\ref{FDM2}.

\begin{figure*}[htbp]
	\centering
	\includegraphics[width=1.0\textwidth]{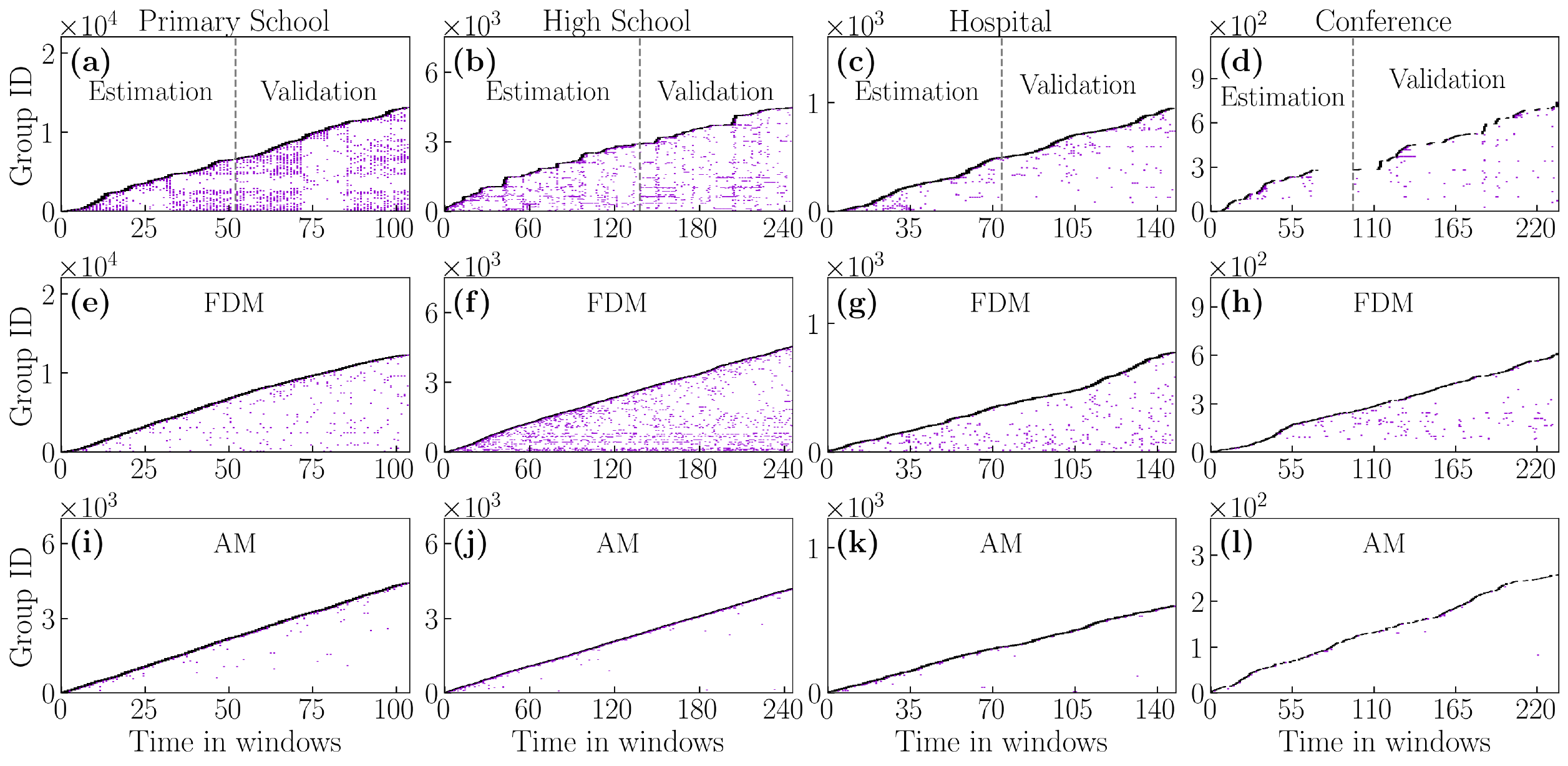}
	\caption{\textbf{Recurrence of groups in real-world and simulate networks, in which the groups include triangles, tetrahedrons, pentachorons, etc.} 
		(a-d) show the recurrent patterns of groups in different real-world networks. The gray dashed line separates the estimation and validation sections used in the model. (e-h) and (i-l) show the recurrent patterns of groups for the corresponding networks simulated by the FDM and AM models, respectively. In each figure, the purple lines represent the recurrent groups, while the black ones correspond to the first occurrence of a group. We bin the x-axis into 10-minute intervals as a time window and obtain an aggregated network snapshot for each time window. The y-axis shows the group IDs in each time window.}
	\label{unsplit}
\end{figure*}

We show the recurrent patterns of groups in different real networks in Fig.~\ref{unsplit}, where the groups include triangles, tetrahedrons, pentachorons, etc. We find the results are robust.

\begin{figure*}[t]
	\centering
	\includegraphics[width=1.0\textwidth]{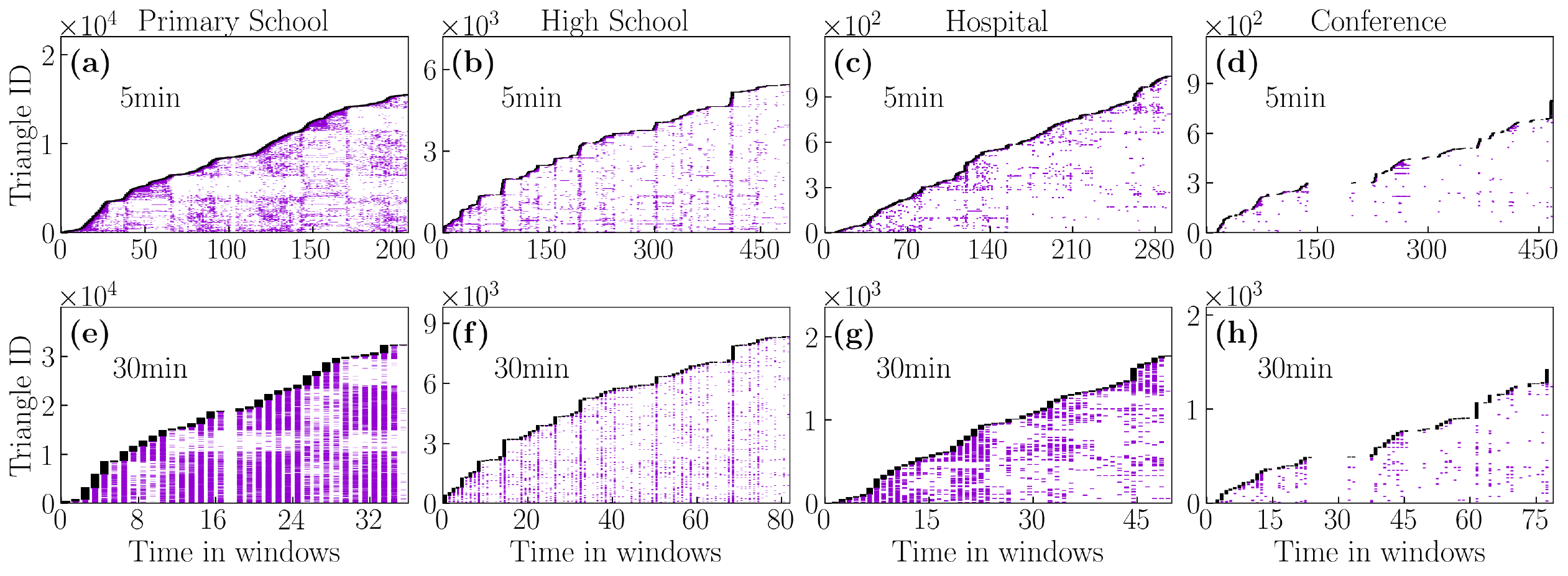}
	\caption{\textbf{The recurrent patterns of full triangles in  aggregate network snapshots with $5$ or $30$ min interval.} The first and second row show the recurrent patterns of full triangles in different real-world networks by aggregating time slots into a snapshot with a $5$-minute and $30$-minute intervals, respectively. Each column corresponds to a dataset.}
	\label{S1}
\end{figure*}

We bin the empirical data nonoverlapping with $5$ and $30$ min interval as a time window and aggregate the networks in each bin to a snapshot in Fig.~\ref{S1}. We observe the same phenomenon about the recurrent patterns of groups in the real-world networks.

\begin{figure*}[htbp]
	\centering
	\includegraphics[width=1.0\textwidth]{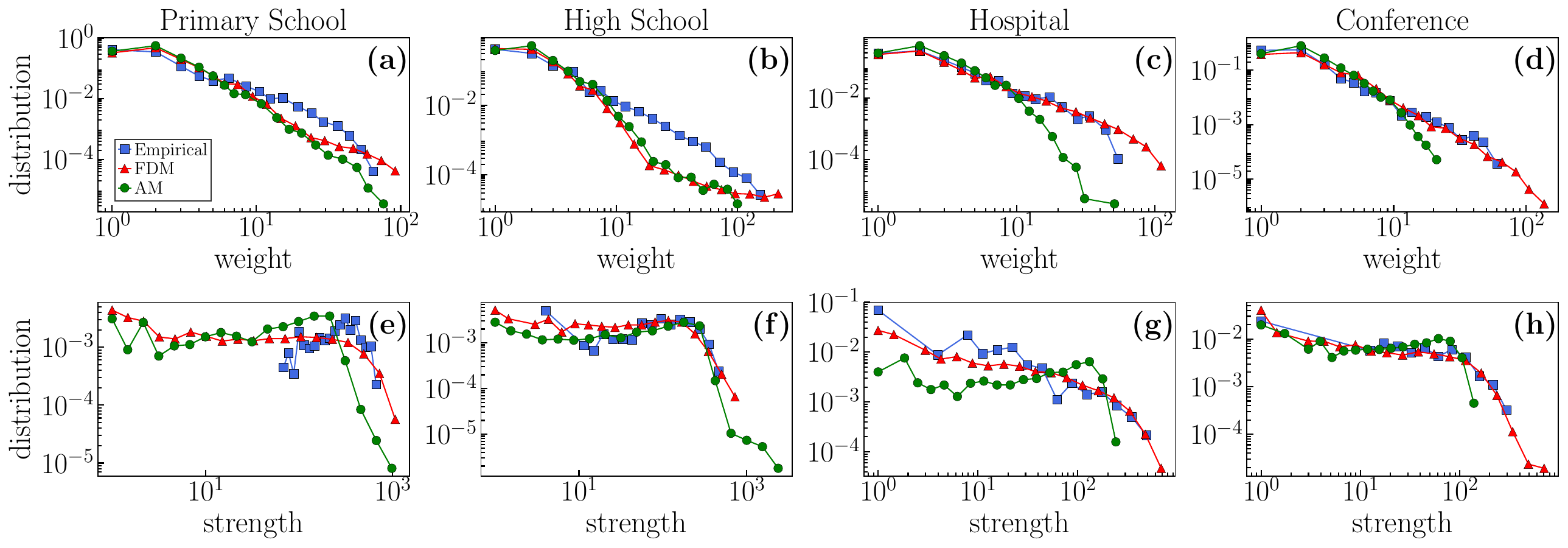}
	\caption{\textbf{Network properties of the real-world datasets and corresponding simulated networks.}
		(a-d) show the weight distribution, where the the weight represents the number of occurrences of each edge in the aggregated network snapshots. (e-h) show the strength distribution, where the strength of a node is the sum of the weights of all edges connected to that node in the aggregated network snapshots. The results represent the averages over 20 networks.}
	\label{S2}
\end{figure*}

We explore additional topological and structural properties of the real-world networks and simulated networks by the FDM and AM models. As shown in Fig.~\ref{S2}, both FDM and AM networks effectively replicate the distributions of edge weights and node strengths of the real-world networks.

\begin{figure*}[htbp]
	\centering
	\includegraphics[width=0.85\textwidth]{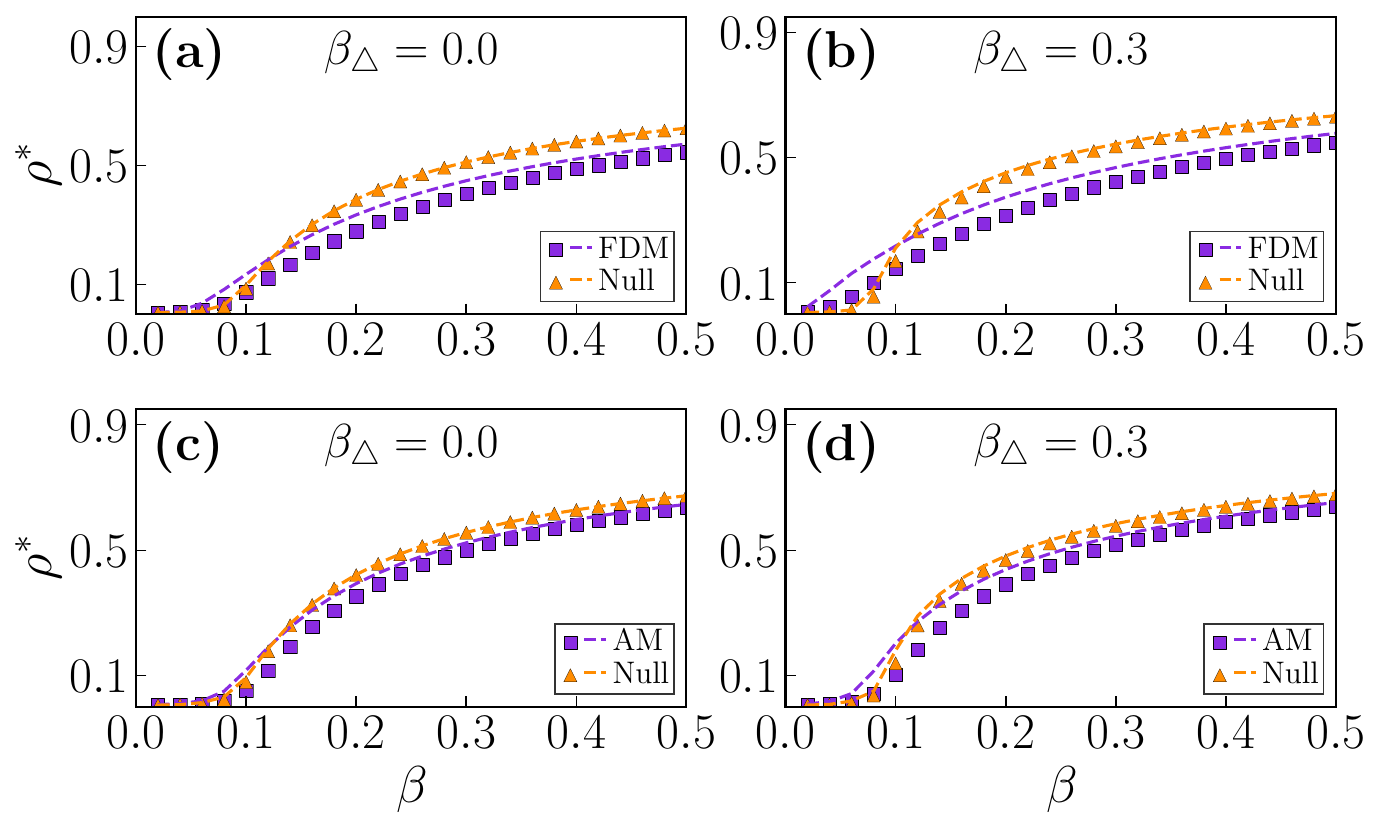}
		\caption{\textbf{Recurrence of full triangles inhibits the information spreading in the FDM and AM models using the parameters in high school.} 
		Each figure shows the $\rho^*$ as a function of $\beta$ for simulated networks by the FDM or AM model and their corresponding null models. The first and second columns show the spreading range $\rho^*$ without higher-order interactions (i.e., $\beta_{\triangle} = 0.0$) and with higher-order interactions (i.e., $\beta_{\triangle} = 0.3$), respectively. Symbols and dashed lines in the figure represent the simulation and MMCA results, respectively. Other parameters are same as Fig.~6 in main text. }
	\label{high_school}
\end{figure*}

\begin{figure*}[htbp]
	\centering
	\includegraphics[width=0.85\textwidth]{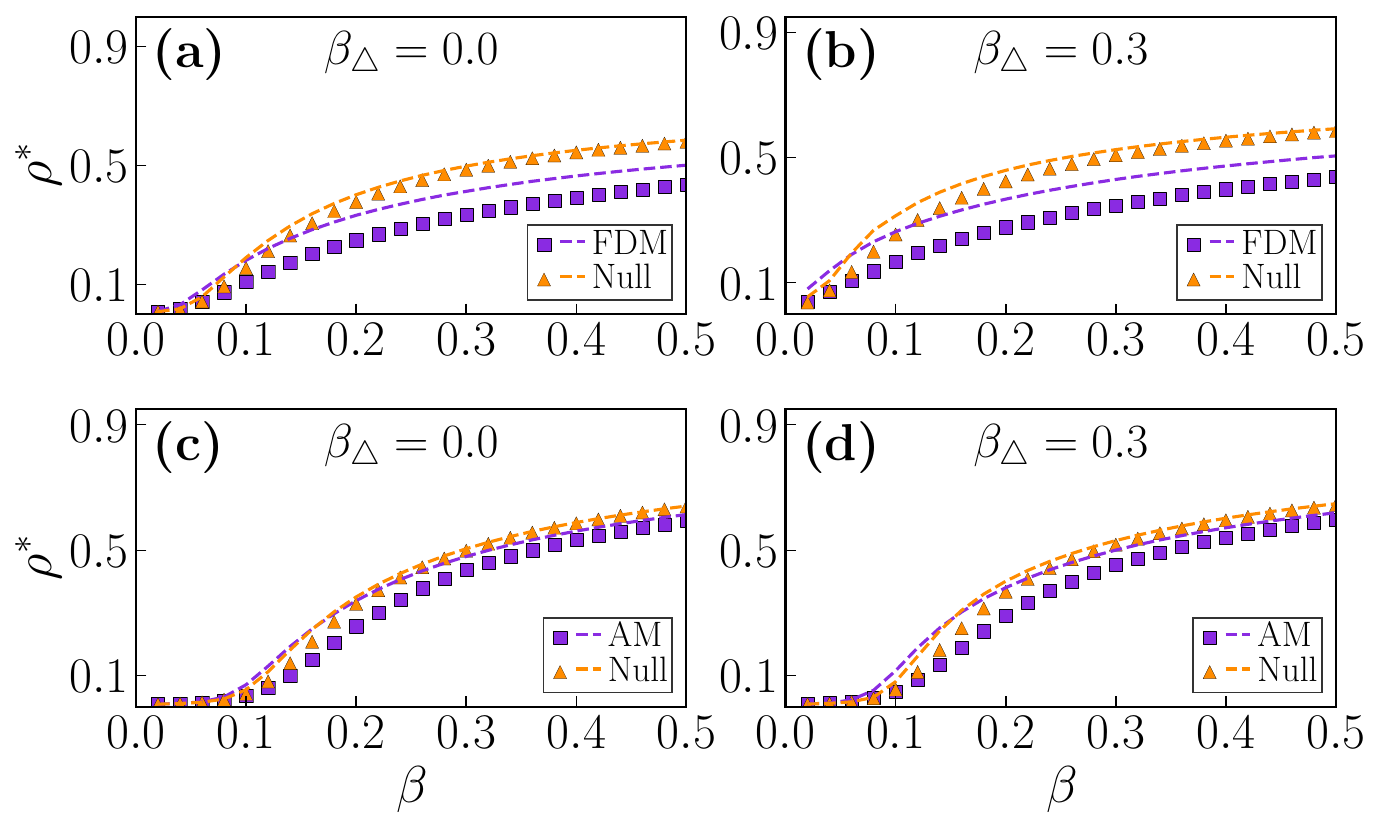}
		\caption{\textbf{Recurrence of full triangles inhibits the information spreading in the FDM and AM models using the parameters in hospital.} 
		Each figure shows the $\rho^*$ as a function of $\beta$ for simulated networks by the FDM or AM model and their corresponding null models. The first and second columns show the spreading range $\rho^*$ without higher-order interactions (i.e., $\beta_{\triangle} = 0.0$) and with higher-order interactions (i.e., $\beta_{\triangle} = 0.3$), respectively. Symbols and dashed lines in the figure represent the simulation and MMCA results, respectively. Other parameters are same as Fig.~6 in main text. }
	\label{hospital}
\end{figure*}

\begin{figure*}[htbp]
	\centering
	\includegraphics[width=0.85\textwidth]{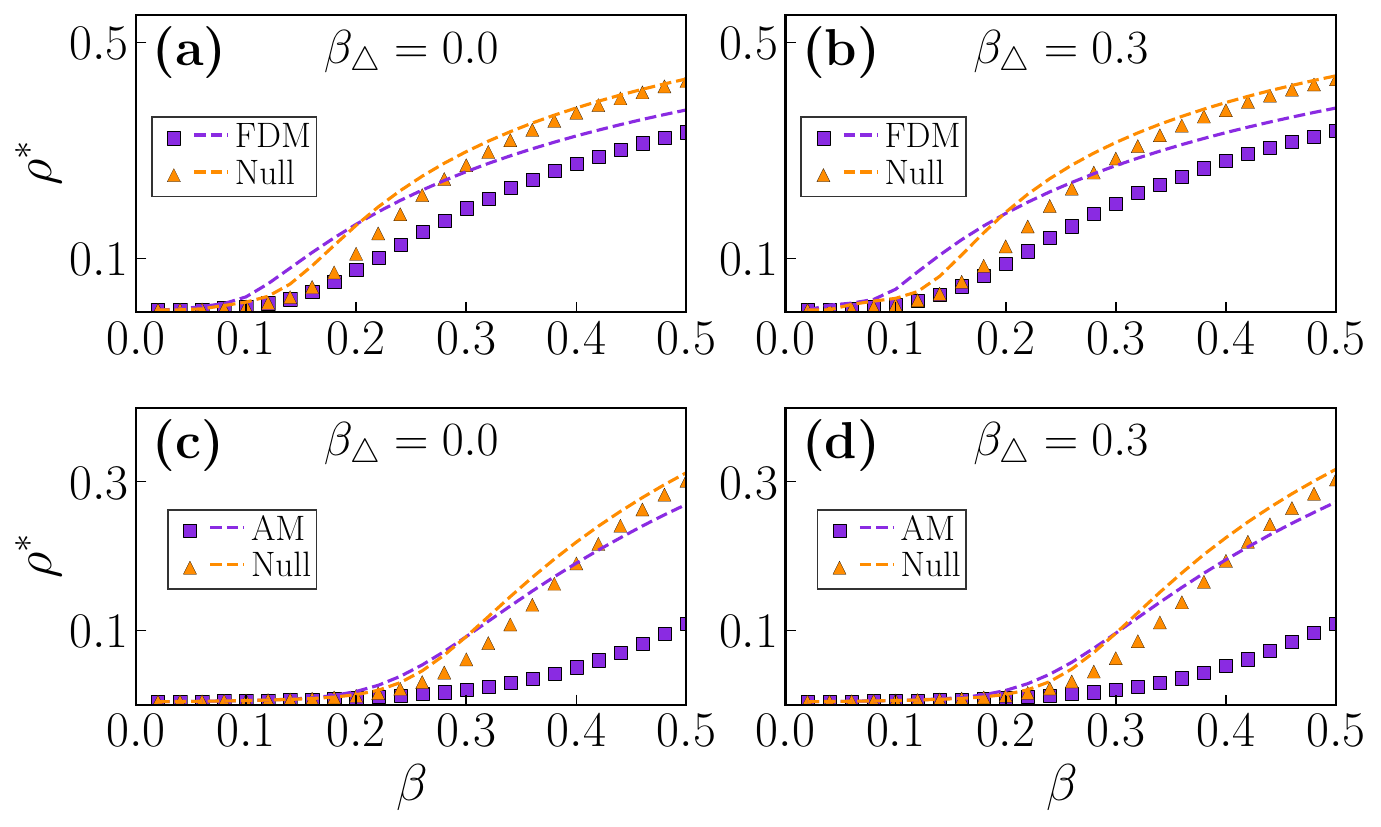}
	\caption{\textbf{Recurrence of full triangles inhibits the information spreading in the FDM and AM models using the parameters in conference.} 
		Each figure shows the $\rho^*$ as a function of $\beta$ for simulated networks by the FDM or AM model and their corresponding null models. The first and second columns show the spreading range $\rho^*$ without higher-order interactions (i.e., $\beta_{\triangle} = 0.0$) and with higher-order interactions (i.e., $\beta_{\triangle} = 0.3$), respectively. Symbols and dashed lines in the figure represent the simulation and MMCA results, respectively. Other parameters are same as Fig.~6 in main text. }
	\label{conference}
\end{figure*}

We generate the simulated networks with the FDM and AM models using the parameters at different datasets and verify the recurrent groups' impact on information spreading. We also generate the corresponding null model for each simulated network and perform the information spreading dynamics on top of them. The results in Figs.~\ref{high_school}-\ref{conference} show a very similar phenomenon, which confirms that the recurrence of groups inhibits the information spreading, and the higher-order interactions will make this phenomenon more pronounced. 

\begingroup
\renewcommand{\thd}[1]{\multicolumn{1}{l}{#1}}
\begin{table}[htbp] 
	\label{tab:S2}	
	\centering
	\caption{Another set of parameters for FDM model that can reproduce similar results as in main text. $N$ is the number of individuals to simulate. The parameter $T$ denotes the time slots to be simulated, corresponding to the time slots in the observation period. $L$ is the square's side length that defines the boundaries of the two-dimensional Euclidean space. $\mu_{1}$, $F_{0}$, and $\mu_{2}$ are the FDM parameters used for simulating each empirical network.} 
	\begin{tabular}{*{7}{l}}
		\toprule
		\thd{Data set}   & \thd{$N$} & \thd{$T$} & \thd{$L$} & \thd{$\mu_{1}$} & \thd{$F_{0}$} & \thd{$\mu_{2}$} \\
		\midrule
		Primary School          & 242                    &1555                           & 62                      & 0.30                  & 0.16                      & 0.71                                           \\			
		High School             & 327                    &4137                               & 96                      & 1.80                  & 0.37                      & 0.17                         \\
		Hospital                & 70                      &2200                            & 126                     & 0.88                  & 0.10                      & 1.12                        \\
		Conference              & 113                     &2874                             & 114                     & 2.45                  & 0.04                      & 1.27                       \\
		\bottomrule
	\end{tabular}
\end{table}
\endgroup

\begingroup
\renewcommand{\thd}[1]{\multicolumn{1}{l}{#1}}
\begin{table}[htbp] 
	\label{tab:S3}	
	\centering
	\caption{Another set of parameters for FDM model that can reproduce similar results as in main text. $N$ is the number of individuals to simulate. The parameter $T$ denotes the time slots to be simulated, corresponding to the time slots in the observation period. $L$ is the square's side length that defines the boundaries of the two-dimensional Euclidean space. $\mu_{1}$, $F_{0}$, and $\mu_{2}$ are the FDM parameters used for simulating each empirical network.} 
	\begin{tabular}{*{7}{l}}
		\toprule
		\thd{Data set}   & \thd{$N$} & \thd{$T$} & \thd{$L$} & \thd{$\mu_{1}$} & \thd{$F_{0}$} & \thd{$\mu_{2}$} \\
		\midrule
		Primary School          & 242                    &1555                           & 64                      & 0.45                  & 0.13                      & 0.82                                           \\			
		High School             & 327                    &4137                               & 94                      & 2.20                  & 0.34                      & 0.16                         \\
		Hospital                & 70                      &2200                            & 124                     & 0.76                  & 0.11                      & 1.06                        \\
		Conference              & 113                     &2874                             & 153                     & 2.05                  & 0.04                      & 1.18                       \\
		\bottomrule
	\end{tabular}
\end{table}
\endgroup


\bibliography{reference}